\documentclass[11pt]{article}
\usepackage{moriond,epsfig}
\usepackage{hyperref}
\pdfoutput=1

\newcommand{\jpsi}{{\rm{J}/\psi}}

\newcommand{\pp}{{p+p}}
\newcommand{\pa}{{p+{\rm A}}}
\newcommand{\da}{{d+{\rm A}}}
\newcommand{\ncol}{{N_{\rm col}}}
\newcommand{\npart}{{N_{\rm part}}}
\renewcommand{\aa}{{\rm{A+A}}}

\newcommand{\dau}{{d+{\rm Au}}}
\newcommand{\auau}{{\rm{Au+Au}}}
\newcommand{\cucu}{{\rm{Cu+Cu}}}

\newcommand{\rda}{{R_{d{\rm A}}}}
\newcommand{\rdau}{{R_{d{\rm Au}}}}
\newcommand{\rcp}{{R_{\rm CP}}}
\newcommand{\raa}{{R_{\rm AA}}}

\begin{document}
\vspace*{4cm}
\title{Heavy Quarkonia Production in Relativistic {\em d}+A and A+A collisions at RHIC, as Measured by the PHENIX Experiment}

\author{H. Pereira Da Costa, for the PHENIX collaboration}

\address{IRFU/Sphn, CEA Saclay, \\F-91191, Gif-sur-Yvette, France}

\maketitle
\abstracts{
This contribution presents the latest results on heavy quarkonia ($\jpsi$ and $\Upsilon$) production in $\da$ and $\aa$ collisions at RHIC at a center of mass energy per nucleon-nucleon collision $\sqrt{s_{\rm NN}}=200$~GeV, measured by the PHENIX experiment, as well as their implications in terms of understanding cold nuclear matter effects and the possible formation of a quark gluon plasma in central $\auau$ collisions.}

Heavy quarkonia have long been considered a favored probe to study the formation of a Quark Gluon Plasma (QGP) in heavy ion collisions for the following reasons: 1) due to their large masses, they are hard probes produced during the first instants of the collision via hard scattering of partons (dominantly gluons, at RHIC); 2) their large binding energy makes them harder to dissociate via interactions with the surrounding hadrons; 3) their production was predicted to be affected by the formation of a QGP, provided that the temperature is high enough, via a color screening mechanism similar to the Debye screening in QED~\cite{Matsui}.

However their production is also affected by other mechanism that do not necessitate the formation of a QGP, and are therefore referred to as Cold Nuclear Matter (CNM) effects. Such effects include: 1) the modification of the Parton Distribution Functions (PDF) in the nuclei, which relates to the fact that the number of partons at a given momentum fraction $x$ inside a nucleon is different whether the nucleon is isolated or inside a nucleus, thus affecting the momentum distribution of quarkonia formed via the hard scattering of these partons~\cite{Vogt,EKS,EPS09} (note that a qualitatively similar effect exists at small $x$ in the so-called {\em gluon saturation} region, which can be described in the Color Glass Condensate framework~\cite{CGC}); 2) the energy loss (via inelastic scattering) of the projectile parton inside the target nucleus before the hard scattering leading to the quarkonia formation, denoted as {\em initial state energy loss}~\cite{ELoss}; 3) the broadening of the quarkonia transverse momentum distribution due to elastic scattering of its parent partons inside the nucleus, denoted as Cronin effect~\cite{Cronin}; 4) the dissociation of the quarkonia bound state (or its precursors) by scattering off the surrounding hadrons~\cite{Vogt}. These effects can be quantified by studying the production of heavy quarkonia in $\pa$ (or rather $\da$, at RHIC) collisions.

Two quantities are used in this presentation to characterize changes to heavy quarkonia production in $\da$ and $\aa$ collisions: 
\begin{itemize}
\item the nuclear modification factor $\rda$ (or $\raa$), obtained by forming the ratio of the heavy quarkonia yield measured in $\da$ ($\aa$) collisions to the yield measured in $\pp$ collisions, normalized by the number of binary ($\pp$) collisions, $\ncol$, corresponding to one $\da$ collision. This factor does not depend on the heavy quarkonia production mechanism (provided that this mechanism is a hard process), and is equal to unity if $\da$ collisions can effectively be considered as an incoherent superposition of independent $\pp$ collisions, as far as heavy quarkonia production is concerned. $\rda$ can be evaluated in bins of the heavy quarkonia rapidity ($y$) and transverse momentum, but also as a function of the collision centrality, which is related to the distance between the target and the projectile nuclei centers (also called {\em impact parameter}, $b$). Central collisions correspond to small values of $b$, whereas peripheral collisions correspond to large values of $b$. Note that $b$ cannot be measured directly, and that depending on how the centrality is actually measured, its correlation to $b$ can be rather lose (especially in the $\da$ case).
\item the central to peripheral ratio $\rcp$, obtained by forming the ratio of the heavy quarkonia yields between central collisions (for which maximum modifications are expected) and peripheral collisions (which should be more similar to the $\pp$ case), properly normalized by the corresponding $\ncol$ values. The advantage of $\rcp$ over $\rda$ is that most experimental systematic uncertainties are canceled, since the same data set can be used for the numerator and the denominator. On the other hand, the quantitative interpretation of $\rcp$ deviations from unity is more ambiguous due to the fact that peripheral collisions cannot be strictly considered identical to $\pp$.
\end{itemize}

\begin{figure}[htb]
\begin{tabular}{cc}
\parbox{5.5cm}{\includegraphics[height=5.8cm]{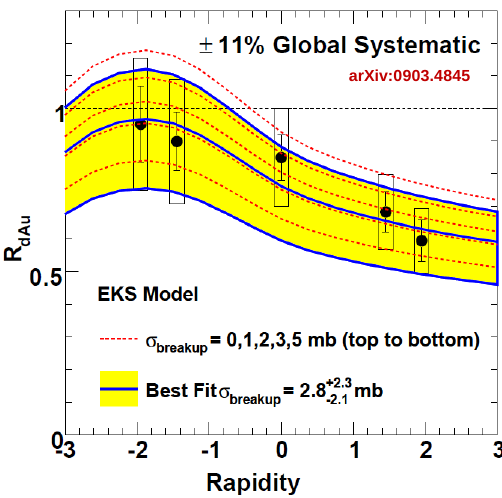}}&
\parbox{12cm}{\includegraphics[height=5.2cm]{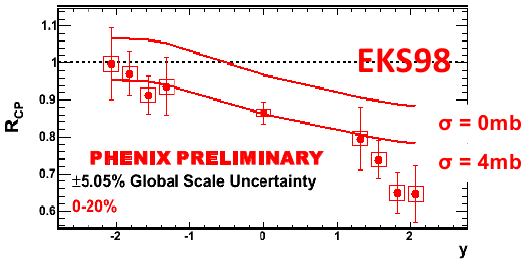}}\\
\end{tabular}
\caption{\label{rda}Left: $\jpsi$ $\rdau$ as a function of rapidity using 2003 $\dau$ minimum bias collisions. Right: $\jpsi$ $\rcp$ as a function of rapidity using 2008 $\dau$ central ($0-20$~\%) and peripheral ($60-88$~\%) collisions. Solid lines correspond to calculations from R. Vogt based on the EKS shadowing model convoluted to different values for the $\jpsi$ breakup cross-section $\sigma_{\rm breakup}$.}
\end{figure}

Figure~\ref{rda} (left) shows the published $\jpsi$ $\rdau$ measured by PHENIX using RHIC 2003 $\dau$ minimum bias data as a function of the $\jpsi$ rapidity~\cite{phenix2003}. $\jpsi$ produced at negative rapidity (the gold going side) originate from gluons of rather large $x$ in the gold nuclei, with $x$ being the fraction of the nucleon momentum carried by the gluon. On the contrary, $\jpsi$ produced at positive rapidity originate from gluons of rather small $x$ in the gold nuclei. A $\jpsi$ suppression is observed at positive rapidity, which is consistent with gluon shadowing (and/or saturation) in the small $x$ regime inside the gold nuclei. The various curves shown on the figure correspond to model calculations that include a parametrization of the gluon shadowing (here EKS~\cite{EKS}) together with a $\jpsi$ break-up cross-section $\sigma_{\rm breakup}$ that varies from 0 to 5~mb. These curves can be used to fit the data and extract the most probable value for $\sigma_{\rm breakup}$. The shape obtained using this most probable value describes the data reasonably well, within the large statistical and systematic uncertainties. 

Figure~\ref{rda} (right) shows the preliminary $\jpsi$ $\rcp$ measured by PHENIX using RHIC 2008 $\dau$ data. This data set corresponds to an integrated luminosity of $\sim 55$~nb$^{-1}$, which represents an increase of about a factor 40 with respect to the 2003 data set, thus allowing to have more experimental points as a function of rapidity and use several centrality bins. The curves shown on the figure have been obtained using a similar approach to that of the left panel. Due to the reduced uncertainties and the larger number of experimental points, it appears that this approach (shadowing + $\sigma_{\rm breakup}$) is unable to properly describe the data, especially at the most forward rapidity, and that additional effects, such as initial state energy loss, must be accounted for in the calculation. Alternatively, one can parametrize the difference between the shadowing calculation (obtained with $\sigma_{\rm breakup}=0$) and the data by a phenomenological rapidity dependent cross-section. Such a cross-section naturally increases for positive rapidity. A qualitatively similar trend has also been observed for lower energy $\jpsi$ measurements~\cite{lourenzo}. 

\begin{figure}[htb]
\begin{tabular}{cc}
\parbox{7cm}{\includegraphics[height=7.5cm]{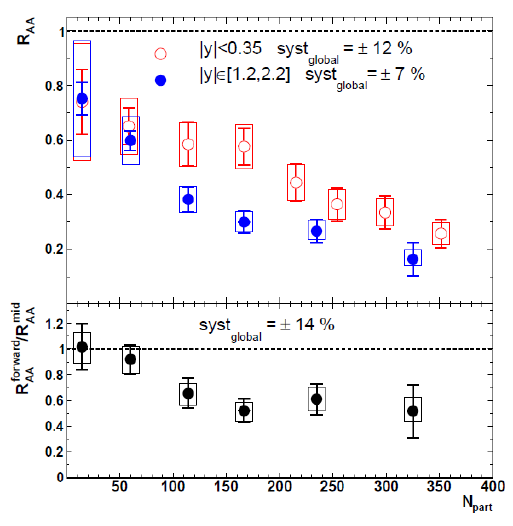}}&
\parbox{6cm}{\includegraphics[height=6cm]{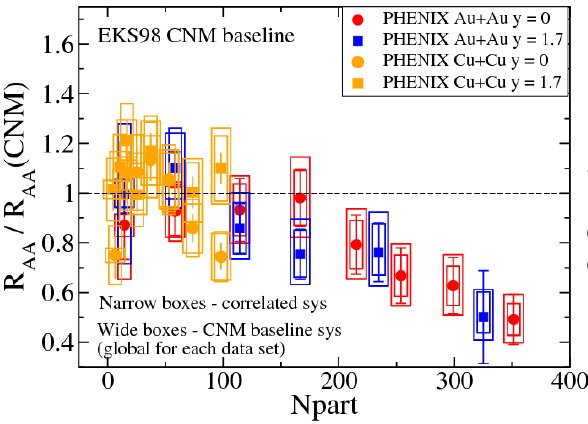}}\\
\end{tabular}
\caption{\label{raa}Left: $\jpsi$ $\raa$ as a function of centrality in 2004 $\auau$ collisions, for two rapidity bins. Right: $\jpsi$ $\raa$ divided by extrapolated cold nuclear matter effects based on 2008 $\dau$ data set and using different values for $\sigma_{\rm breakup}$ as a function of rapidity.}
\end{figure}

A significant $\jpsi$ suppression has also been measured in $\auau$ collisions as illustrated as a function of centrality (here $\npart$, the average number of nucleons that participate to $\auau$ collisions at a given centrality) in figure~\ref{raa} (left) for two rapidity regions~\cite{phenix2004}. It appears that the suppression is larger at forward rapidity than at mid-rapidity. It is important to quantify which fraction of this suppression is due to CNM effects alone, in order to single out the effect of the possible formation of a QGP. This is achieved by using a parametrization of the observed $\dau$ suppression (using a shadowing model and a rapidity dependent $\sigma_{\rm breakup}$) and extrapolating it to the $\auau$ case. The measured $\raa$ can then be divided by this CNM-only expected suppression, as shown in figure~\ref{raa} (right)~\cite{frawley}. A significant $\jpsi$ suppression remains for central enough collisions (large values of $\npart$) even after CNM effects have been removed. This indicates that additional effects (for instance due to the formation of a QGP) must be accounted for to be able to describe the $\jpsi$ suppression observed in central $\auau$ collisions. Besides, this remaining suppression is now similar between the two rapidity domains, meaning that the differences observed in figure~\ref{raa} (left) can be entirely accounted for by CNM effects. Also shown on the figure are the results obtained in $\cucu$ collisions~\cite{phenix2005} and for which no significant suppression beyond CNM effects is observed. 

\begin{figure}[htb]
\begin{tabular}{cc}
\parbox{7cm}{\includegraphics[height=4.8cm]{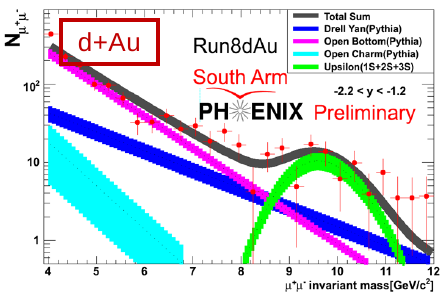}}&
\parbox{6.5cm}{\includegraphics[height=5cm]{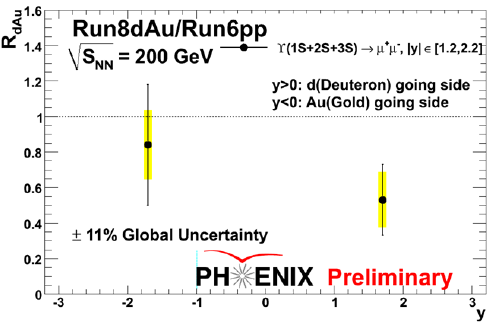}}\\
\end{tabular}
\caption{\label{upsilon}Left: di-muon invariant mass distribution at the $\Upsilon$ mass measured using 2008 $\dau$ collisions. Coloured area correspond to the different contribution to the di-muon spectra. Right: $\Upsilon$ $\rda$ as a function of rapidity.}
\end{figure}

Finally, $\Upsilon$ resonances have been measured by PHENIX in both $\pp$ (using 2006 data set) and $\dau$ collisions (using the 2008 data set), allowing one to form the first $\Upsilon$ $\rda$ at this energy. Figure~\ref{upsilon} (left) shows the di-muon invariant mass distribution at the $\Upsilon$ mass, measured in one of PHENIX muon arms in $\dau$ collisions. The various coloured bands correspond to the different contributions to this spectra, and are fitted to the data in order to evaluate the number of produced $\Upsilon$ resonances. Figure~\ref{upsilon} (right) shows the $\Upsilon$ $\rda$ obtained by combining these data to yields measured in $\pp$ collisions. A suppression is observed at forward rapidity, similarly to the $\jpsi$ case, which has yet to be compared to theoretical calculations. 

$\Upsilon$ resonances have also been measured in $\auau$ collisions at mid-rapidity by the PHENIX experiment. However, due to the limited statistics, only a 90~\% Confidence Level (CL) could be evaluated so far for $\raa$ and the contributions to the di-lepton invariant mass spectrum from other (background) sources in the $\Upsilon$ mass region have not been accounted for. The resulting value is $\raa < 0.64$ at 90~\% CL for all di-leptons produced at mid-rapidity ($|y|<0.35$) with mass $M\in[8.5,11.5]$~GeV/c$^2$.

To summarize, a significant $\jpsi$ suppression has been observed by PHENIX in both 2003 and 2008 $\dau$ data sets at positive rapidity (in the gluon shadowing/saturation region for the gold nuclei), with an increased statistics of about a factor 40 for the 2008 data. This suppression can be attributed to CNM effects, but cannot be properly described (for the 2008 data) with a simple model that considers only PDF modifications and a (rapidity independent) $\jpsi$ break-up cross section. Phenomenologically parametrized CNM effects can be extrapolated to $\auau$ collisions and compared to the $\jpsi$ suppression measured in these conditions. It appears that an additional suppression (not attributed to CNM effects) remains for central $\auau$ collisions, which is largely rapidity independent. Concerning the $\Upsilon$ resonance, a suppression similar to that of the $\jpsi$ has also been observed for the first time in $\dau$ collisions and a 90~\% confidence level for high mass di-lepton pairs (dominantly $\Upsilon$ mesons) $\raa$ has measured.

\end{document}